\documentclass[runningheads]{llncs}
\usepackage{graphicx}
\usepackage{authblk}
\usepackage{tikz}
\usepackage{comment}
\usepackage{amsmath,amssymb} 
\usepackage{color}

\usepackage{booktabs}
\usepackage[noend,ruled]{algorithm2e}
\usepackage{wrapfig}

\usepackage[accsupp]{axessibility}  

\newcommand{\redbf}[1]{{\textbf{\color{black}{#1}}}} 
\newcommand{\blueud}[1]{{\underline{\color{black}{#1}}}} %

\usepackage{hyperref}
\hypersetup{
	colorlinks=true,
	linkcolor=red,
	filecolor=red,      
	urlcolor=blue,
}

\makeatletter
\DeclareRobustCommand\onedot{\futurelet\@let@token\@onedot}
\def\@onedot{\ifx\@let@token.\else.\null\fi\xspace}

\def\eg{\emph{e.g}\onedot} 
\def\ie{\emph{i.e}\onedot} 
 
\def\etc{\emph{etc}\onedot}

\def\etal{\emph{et al}\onedot}
\makeatother

\begin{document}
\pagestyle{headings}
\mainmatter
\def\ECCVSubNumber{1129}  

\makeatletter
\renewcommand\AB@affilsepx{, \protect\Affilfont}
\makeatother

\title{Auto-FedRL: Federated Hyperparameter Optimization for Multi-institutional Medical Image Segmentation} 

\institute{ }
\titlerunning{Auto-FedRL: Federated Hyperparameter Optimization}
%
\author{Pengfei Guo\thanks{Work done during an internship at NVIDIA. NVFlare~\cite{nvflare} implementation of this work is available at \url{https://nvidia.github.io/NVFlare/research/auto-fed-rl}.}\inst{1}\and
Dong Yang\inst{2} \and
Ali Hatamizadeh\inst{2} \and
An Xu\inst{3} \and
Ziyue Xu\inst{2} \and
Wenqi Li\inst{2} \and
Can Zhao\inst{2} \and
Daguang Xu\inst{2} \and
Stephanie Harmon\inst{4} \and
Evrim Turkbey\inst{5} \and
Baris Turkbey\inst{4} \and
Bradford Wood\inst{5} \and
Francesca Patella\inst{6} \and
Elvira Stellato\inst{7} \and
Gianpaolo Carrafiello\inst{7} \and
Vishal M. Patel\inst{1} \and
Holger R. Roth\inst{2}
}
\authorrunning{P. Guo et al.}
\institute{Johns Hopkins University \and NVIDIA \and
University of Pittsburgh \and
National Cancer Institute \and
National Institutes of Health \and
ASST Santi Paolo e Carlo \and
University of Milan
}

\maketitle

\begin{abstract}
Federated learning (FL) is a distributed machine learning technique that enables collaborative model training while avoiding explicit data sharing. The inherent privacy-preserving property of FL algorithms makes them especially attractive to the medical field. However, in case of heterogeneous client data distributions, standard FL methods are unstable and require intensive hyperparameter tuning to achieve optimal performance. Conventional hyperparameter optimization algorithms are impractical in real-world FL applications as they involve numerous training trials, which are often not affordable with limited compute budgets. In this work, we propose an efficient reinforcement learning~(RL)-based federated hyperparameter optimization algorithm, termed Auto-FedRL, in which an online RL agent can dynamically adjust hyperparameters of each client based on the current training progress. Extensive experiments are conducted to investigate different search strategies and RL agents. The effectiveness of the proposed method is validated on a heterogeneous data split of the CIFAR-10 dataset as well as two real-world medical image segmentation datasets for COVID-19 lesion segmentation in chest CT and pancreas segmentation in abdominal CT.

\end{abstract}

\section{Introduction}\label{intro}
A large amount of data is needed to train robust and generalizable machine learning models. A single institution often does not have enough data to train such models effectively.
Meanwhile, there are emerging regulatory and privacy concerns about the data sharing and management~\cite{kaissis2021end,invertg}.
Federated Learning (FL)~\cite{fedavg} mitigates such concerns as it leverages data from different clients or institutions to collaboratively train a global model while allowing the data owners to  control their private datasets. 
Unlike the conventional centralized training, FL algorithms open the potential for multi-institutional collaborations in a privacy-preserving manner~\cite{yang2019federated}. 
This multi-institutional collaboration scenario often refers to \emph{cross-silo} FL~\cite{kairouz2019advances} and is the main focus of this paper. In this FL setting, clients are autonomous data owners, such as medical institutions storing patients' data, and collaboratively train a general model to overcome the data scarcity issue and privacy concerns~\cite{yang2019federated}. This makes \emph{cross-silo} FL applications especially attractive to the healthcare sector~\cite{rieke2020future,guo4049653learning}. Several methods have already been proposed to leverage FL for multi-institutional collaborations in digital healthcare~\cite{autofedavg,flmr,sheller2020federated,roth2020federated}.

The most recently introduced FL frameworks~\cite{autofedavg,flmr,intel,fedprox} are variations of the Federated Averaging (FedAvg)~\cite{fedavg} algorithm. The training process of FedAvg consists of the following steps: (i) clients perform local training and upload model parameters to the server. (ii) The server carries out the averaging aggregation over the received parameters from clients and broadcasts aggregated parameters to clients. (iii) Clients update local models and evaluate its performance. After sufficient communication rounds between clients and the server, a global model can be obtained. The design of FedAvg is based on standard Stochastic Gradient Descent (SGD) learning with the assumption that data is uniformly distributed across clients~\cite{fedavg}. However, in real-world applications, one has to deal with underlying unknown data distributions that are likely not independent and identically distributed
(non-iid). The heterogeneity of data distributions has been identified as a critical problem that causes the local models to diverge during training and consequently sub-optimal performance of the trained global model~\cite{fedbn,intel,fedprox}.

To achieve the required performance, the proper tuning of hyperparameters (\eg, the learning rate, the number of local iterations, aggregation weights, \etc) plays a critical role for the success of FL~\cite{intel,autofedavg}. \cite{flconvergence} shows that the learning rate decay is a necessary condition of the convergence for FL on non-iid data. While the general hyperparameter optimization has been intensively studied~\cite{randomsearch,bayesian1,bayesian2}, the unique setting of FL makes federated hyperparameters optimization especially difficult~\cite{weightsharing}. Reinforcement learning (RL) provides a promising solution to approach this complex optimization problem. Compared to other methods for finding the optimal hyperparameters, RL-based methods do not require the prior knowledge of the complicated underlying system dynamics~\cite{rladvantage}. Thus, federated hyperparameter optimization can be reduced to defining appropriate reward metrics, search space, and RL agents.

In this paper, we aim to make the automated hyperparameter optimization applicable in realistic FL applications. An online RL algorithm is proposed to dynamically tune hyperparameters during a single trial. Specifically, the proposed Auto-FedRL formulates hyperparameter optimization as a task of \begin{wrapfigure}{r}{0.4\textwidth}
  \begin{center}
    \includegraphics[width=0.4\textwidth]{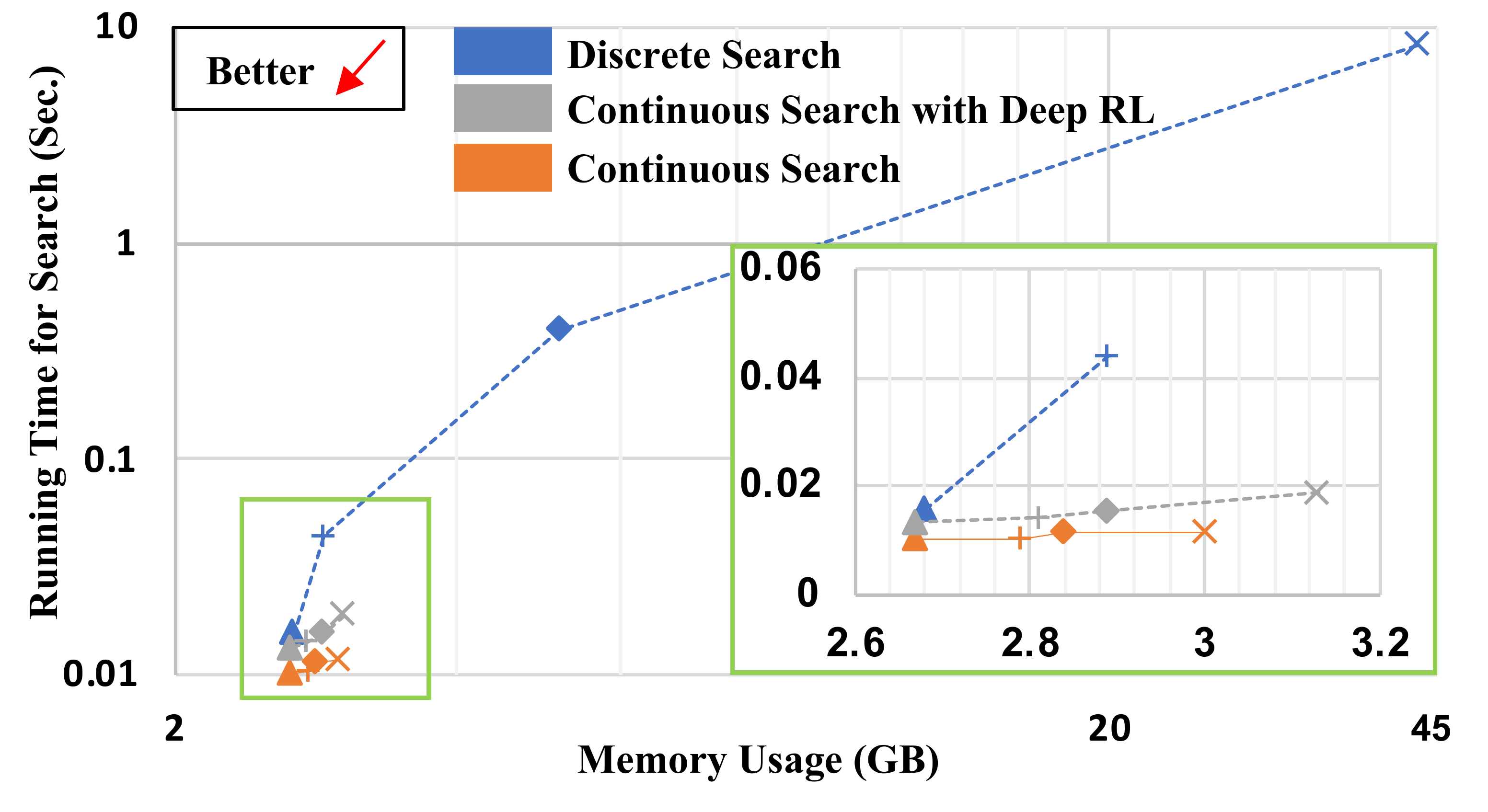}
	\caption{The computational details of different search strategies under the same setting on CIFAR-10 when the number of clients equals to 2 ($\bigtriangleup$), 4 ($+$), 6 ($\Diamond$), and 8 ($\times$). The green box shows the zoomed-in region.}\label{fig:cd}
  \end{center}
\end{wrapfigure}discovering optimal policies for the RL agent. Auto-FedRL can dynamically adjust hyperparameters at each communication round based on relative loss reduction. Without the need for multiple training trails, an online RL agent is introduced to maximize the rewards in small intervals, rather than the sum of all rewards.
While RL-based hyperparameter optimization method has been explored in~\cite{intel}, our experiments show that the prior work has several deficiencies impeding its practical use in real-world applications.
(\textbf{i}) The discrete action space (\ie, hyperparameter search space) not only leads to limited available actions but also suffers from scalability issues. At each optimization step, the gradient of all possible hyperparameter combinations is retained, which causes high memory consumption and computational inefficiency. 
Therefore, as shown in Fig.~\ref{fig:cd}, the hardware limitation can be reached quickly, when one needs to collaborate with multiple institutions using a large search space. 
To circumvent this challenge, Auto-FedRL can leverage continuous search space. Its memory usage is practically constant as the memory consumption per hyperparameter is negligible and does not explode with increased search space and the number of involved clients. Meanwhile, its computational efficiency is significantly improved compared to discrete search space. (\textbf{ii}) The flexibility of hyperparameter search space is limited. \cite{intel} focuses on a small number of
hyperparameters (\eg, one or two hyperparameters) in less general settings.
In contrast, our method is able to tune a wide range of hyperparameters (\eg, client/server learning rates, the number of local iterations, and the aggregation weight of each client) in a realistic FL setting. It is worth noting that the averaging model aggregation is replaced by a pseudo-gradient optimization~\cite{fedopt} in Auto-FedRL. Thus, we are able to search server-side hyperparameters. 
To this end, we propose a more practical federated hyperparameter optimization framework with notable computational efficiency and flexible search space.

Our main contributions in this work are summarized as follows:
\begin{itemize}
\item A novel federated hyperparameter optimization framework Auto-FedRL is proposed, which enables the dynamic tuning of hyperparameters via a single trial.

\item Auto-FedRL makes federated hyperparameter optimization more practical in real-world applications by efficiently incorporating continuous search space and the deep RL agent to tune a wide range of hyperparameters.

\item Extensive experiments on multiple datasets show the superior performance and notable computational efficiency of our methods over existing FL baselines.
\end{itemize}

\section{Related Works}
%
%
\noindent\textbf{Federated Learning on Heterogeneous Data. } The heterogeneous data distribution across clients impedes the real-world deployment of FL applications and draws emerging attentions. Several methods~\cite{noniid1,noniid2,flconvergence,noniid4,noniid5,xu2022closing,mei2022escaping} have been proposed to address this issue. FedOpt~\cite{fedopt} introduced the adaptive federated optimization, which formulated a more flexible FL optimization framework but also introduced more hyperparameters, such as the server learning rate and server-side optimizers. FedProx~\cite{fedprox} and Agnostic Federated Learning (AFL)~\cite{afl} are variants of FedAvg~\cite{fedavg} which attempted to address the learning bias issue of the global models for local clients by imposing additional regularization terms. FedDyn~\cite{feddyn} was proposed to address the problem that the
minima of the local-device level loss are inconsistent with those of the
global loss by introducing a dynamic regularizer for each device. Those works demonstrated good theoretical analysis but are evaluated only on manually created toy datasets. 
Recently, 
FL-MRCM~\cite{flmr} was proposed to address the domain shift issue among different clients by aligning the distribution of latent features between the source domain and the target domain. Although those methods~\cite{li2020multi,flmr} achieved promising results in overcoming domain shift in the multi-institutional collaboration, directly sharing latent features between clients increased privacy concerns.

\noindent\textbf{Conventional Hyperparameter Optimization. } Grid and random search~\cite{randomsearch} can perform automated hyperparameter tuning but require long running time due to often exploring unpromising regions of the search space. While advanced random search~\cite{adrandomsearch} and Bayesian optimization-based search methods~\cite{bayesian1,bayesian2} require fewer iterations, several training trails are required to evaluate the fitness of hyperparameter configurations. Repeating the training process multiple times is impractical in the FL setting, especially for deep learning models, due to the limited communication and compute resources in real-world FL setups.

\noindent\textbf{Federated Hyperparameter Optimization. } Auto-FedAvg~\cite{autofedavg} is a recent automated search method, which only is compatible with differentiable hyperparameters and focuses on searching client aggregation weights. The method proposed in~\cite{intel} is the most relevant to our work. However, as discussed in the previous section, it suffers from limited practicability and flexibility of search space in real-world applications. Inspired by the recent hyperparameter search~\cite{h1,h2,h3} and differentiable~\cite{d1,d2}, evolutionary~\cite{e1,e2} and RL-based automated machine learning methods~\cite{rl1,rl2}, we propose an efficient automated approach with flexible search space to discover a wide range of hyperparameters.
\section{Methodology}
In this section, we first introduce the general notations of FL and the adaptive federated optimization that provides the theoretical foundation of tuning FL server-side hyperparameters (Sec.~\ref{sec:fl}). Then, we describe our method in detail (Sec.~\ref{sec:auto-fedrl}), including online RL-based hyperparameter optimization, the discrete/continuous search space, and the deep RL agent. In addition, we provide theoretical analysis to guarantee the convergence of Auto-FedRL in the supplementary material.

\subsection{Federated Learning}\label{sec:fl}
In a FL system, suppose $K$ clients collaboratively train a global model. The goal is to solve the optimization problem as follows:
\setlength{\belowdisplayskip}{0.5pt} \setlength{\belowdisplayshortskip}{0.5pt}
\setlength{\abovedisplayskip}{0.5pt} \setlength{\abovedisplayshortskip}{0.5pt}
\begin{equation} \label{eq:1}
\begin{aligned}
\min\limits_{x\in \mathbb{R}^d}\frac{1}{K}\sum\limits_{k=1}^{K}\mathcal{L}_k(x),
\end{aligned}
\end{equation}
where $\mathcal{L}_k(x) = \mathbb{E}_{z\sim \mathcal{D}_k} [\mathcal{L}_k(x,z)]$ is the loss function of the $k^{\text{th}}$ client. $z \in \mathcal{Z}$, and $\mathcal{D}_k$ represents the data distribution of the $k^{\text{th}}$ client. Commonly, for two different clients $i$ and $j$, $\mathcal{D}_i$ and $\mathcal{D}_j$ can be dissimilar, so that Eq.~\ref{eq:1} can become nonconvex. A widely used method for solving this optimization problem is FedAvg~\cite{fedavg}. At each round, the server broadcasts the global model to each client. Then, all clients conduct local training on their own data and send back the updated model to the server. Finally, the server updates the global model by a weighted average of these local model updates. FedAvg's server update at round $q$ can be formulated as follows:
\begin{equation} \label{eq:2}
\begin{aligned}
\Theta^{q+1} = \sum\limits_{k=1}^{K} \alpha_k  \Theta^{q}_k,
\end{aligned}
\end{equation}
where $\Theta_k^q$ denotes the local model of $k^{\text{th}}$ client and $\alpha_k$ is the corresponding aggregation weight. The update of global model $\Theta^{q+1}$ in Eq.~\ref{eq:2} can be further rewritten as follows:
\begin{equation} \label{eq:3}
\begin{aligned}
\Theta^{q+1} & = \Theta^{q} - \sum\limits_{k=1}^{K} \alpha_k (\Theta^{q}-\Theta^{q}_k)\\
& = \Theta^{q} - \sum\limits_{k=1}^{K} \alpha_k\Delta_k^q \\
& = \Theta^{q} - \Delta^q,
\end{aligned}
\end{equation}
where $\Delta_k^q := \Theta^{q}-\Theta^{q}_k$ and $\Delta^q := \sum\limits_{k=1}^{K} \alpha_k \Delta_k^q$. Therefore, the server update in FedAvg is equivalent to applying optimization to the \emph{pseudo-gradient} $-\Delta^q$ with a learning rate $\gamma = 1$. This general FL optimization formulation refers to adaptive federated optimization~\cite{fedopt}. 
Auto-FedRL utilizes this \emph{pseudo-gradient} update formulation to enable the server-side hyperparameter optimization, such as the server learning rate $\gamma$.

\subsection{Auto-FedRL}\label{sec:auto-fedrl}
\noindent\textbf{Online RL Hyperparameter Optimization.}
The online setting in the targeted task is very challenging since the same actions at different training stages may receive various responses. Several methods~\cite{rl_ns1,rl_ns2,rl_ns3} have been proposed in the literature to deal with such non-stationary problems. However, these methods require multiple training runs, which is usually not affordable in FL settings where clients often have limited computation resources. Typically, a client can run only one training procedure at the same time and the resources for parallelization as would be done in a cluster environment is not available. 
To circumvent the limitations of conventional hyperparameter optimization methods and inspired by previous works~\cite{rl2,intel,rl1}, we introduce an online RL-based approach to directly learn the proper hyperparameters from data at the clients' side during a single training trial. At round $q$, a set of hyperparameters $h^q$ can be sampled from the distribution $P(\mathcal{H}| \psi^q)$. We denote the validation loss of client $k$ at round $q$ as $\mathcal{L}_{\text{val}_{k}}^q$ 
and the hyperparameter loss at round $q$ as
\begin{equation}
\begin{aligned}
\mathcal{L}_h^q=\frac{1}{K}\sum\limits_{k=1}^{K} \mathcal{L}_{\text{val}_{k}}^q.
\end{aligned}
\end{equation}
 The relative loss reduction reward function of the RL agent is defined as follows:
\begin{equation} \label{eq:4}
\begin{aligned}
r^q = \frac{\mathcal{L}_h^q-\mathcal{L}_h^{q+1}}{\mathcal{L}_h^q}.
\end{aligned}
\end{equation}
The goal of the RL agent at round $q$ is to maximize the objective as follows:
\begin{equation} \label{eq:5}
\begin{aligned}
J^q = \mathbb{E}_{P(h^q|\psi^q)} [r^q].
\end{aligned}
\end{equation}
By leveraging the one-sample Monte Carlo estimation technique~\cite{rl}, we can approximate the derivative of $J^q$ as follows:
\begin{equation} \label{eq:6}
\begin{aligned}
\nabla_{\psi^q} J^q = r^q \nabla_{\psi^q} \log (P(h^q| \psi^q)).
\end{aligned}
\end{equation}
To this end, we can evaluate Eq.~\ref{eq:5} and use it to update the condition of hyperparameter distribution $\psi^q$. To formulate an online algorithm, we utilize the averaged rewards in a small interval (``window'') rather than counting the sum of all rewards to update $\psi^q$ as follows:
\begin{equation} \label{eq:7}
\begin{aligned}
&\psi^{q+1} \leftarrow \psi^{q} - \gamma_h \sum\limits_{\tau = q-Z}^{\tau = q} (r^\tau - \hat{\tau}^q) \nabla_{\psi^\tau} \log (P(h^\tau| \psi^\tau)),
\end{aligned}
\end{equation}
where $Z$ is the size of the update window and $\gamma_h$ is the RL agent learning rate. The averaged rewards $\hat{\tau}^q$ in the interval $[q-Z,q]$ are defined as follows:
\begin{equation} \label{eq:8}
\begin{aligned}
\hat{\tau}^q = \frac{1}{Z+1} \sum\limits_{\tau = q-Z}^{\tau = q} r^\tau.
\end{aligned}
\end{equation}

\begin{figure}[tbp]
	\centering
	\includegraphics[
	width=\textwidth]{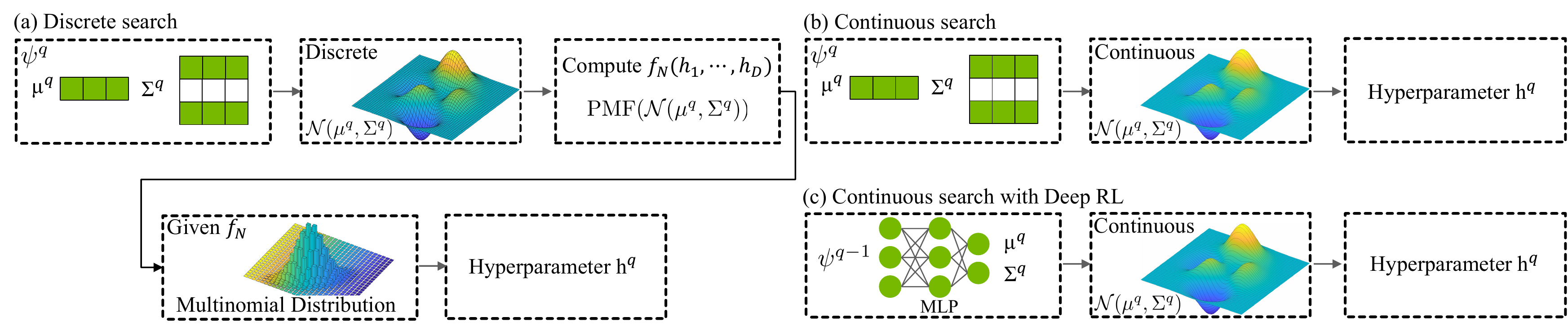}
	\caption{The sampling workflow comparison of different search strategies in the proposed Auto-FedRL. PMF denotes the probability mass function.}\label{fig2}
\end{figure}

\noindent\textbf{Discrete Search. } Selecting the form of hyperparameter distribution $P(\mathcal{H}| \psi)$ is non-trivial, since it determines the available actions in the search space. We denote the proposed method using discrete search (DS) space as Auto-FedRL(DS). Here, $P(\mathcal{H}| \psi)$ is defined by a $D$-dimensional discrete Gaussian distribution, where $D$ denotes the number of searchable hyperparameters. For each hyperparameter, there is a finite set of available selections. Therefore, $\mathcal{H}$ is a grid that consists of all possible combinations of available hyperparameters. A hyperparameter combination $h^q$ at round q is a point on $\mathcal{H}$ as follows:
\begin{equation} \label{eq:9}
\begin{aligned}
P(h^q|\psi^q) = \mathcal{N}(h^q|\mu^q,\Sigma^q),
\end{aligned}
\end{equation}
where $h^q=\{h^q_1,\dots,h_D^q\}$. $\psi^q$ is defined by the mean vector $\mu^q$ and the covariance matrix $\Sigma^q$, which are learnable parameters that the RL agent targets to optimize. To increase the stability of RL training and encourage learning in all directions, different types of predefined hyperparameter selections are normalized to the same scale with zero-mean when constructing the search space. This hyperparameter sampling procedure is presented in Fig.~\ref{fig2} (a).

\noindent\textbf{Continuous Search. } While defining a discrete action space can be more controllable for hyperparameter optimization, as discussed in Section~\ref{intro}, it limits the scalability of the search space. The gradients of all possible hyperparameter combinations are retained in the discrete search during the windowed update as in Eq.~\ref{eq:7}, which requires a large amount of memory. To overcome this issue, we extend Auto-FedRL(DS) to Auto-FedRL(CS), that can utilize a continuous search (CS) space for the RL agent.  
Instead of constructing a gigantic grid that stores all possible hyperparameter combinations, one can directly sample a choice from a continuous multivariate Gaussian distribution $\mathcal{N}(\mu^q,\Sigma^q)$.
It is worth noting that with the expansion of search space, the increase of memory usage of Auto-FedRL(CS) is negligible. A comparison between the hyperparameter sampling workflows in discrete and continuous search are presented in Fig.~\ref{fig2}.
The main difference between DS and CS lies in the sampling process.
In practice, one can adopt the Box–Muller transform~\cite{bmt} for sampling the continuous Gaussian distribution. However, 
as shown in Fig.~\ref{fig2}(a), the sampling for multivariate discrete Gaussian distributions typically involves the following steps: \textbf{(i)} We compute the probabilities of all possible combinations. \textbf{(ii)} Given the probabilities, we draw a choice from the multinomial distribution or alternatively can use the ``inverse CDF'' method~\cite{inversecdf}. In either way, we need to compute the probabilities of all possible hyperparameter combinations for DS, which is not required for CS. Hence, our CS is much more efficient for hyperparameter optimization, as shown in Fig.~\ref{fig:cd}. 

\noindent\textbf{Deep RL Agent. }  An intuitive extension of Auto-FedRL(CS) is to leverage neural networks (NN) as the agent to update the condition of hyperparameter distribution $\psi^{q}$ rather than the direct optimization. A more complicated RL agent design could deal with potentially more complex search spaces~\cite{deeprl}. To investigate the potential of NN-based agent in our setting, we further propose the Auto-FedRL(MLP), which leverages a multilayer perceptron (MLP) as the agent to update the $\psi$. The sampling workflow of Auto-FedRL(MLP) is presented in Fig.~\ref{fig2}(c). The proposed MLP takes the condition of previous hyperparameter distribution $\psi^{q-1}$ as the network's input and predicts the updated $\psi^{q}$. Meanwhile, due to our online setting (\ie limited optimization steps), we have to keep the learnable parameters in MLP small but effective. The detailed network configuration can be found in the supplementary material.
%
%
\begin{figure}[tbp]
\centering
\includegraphics[width=\textwidth]{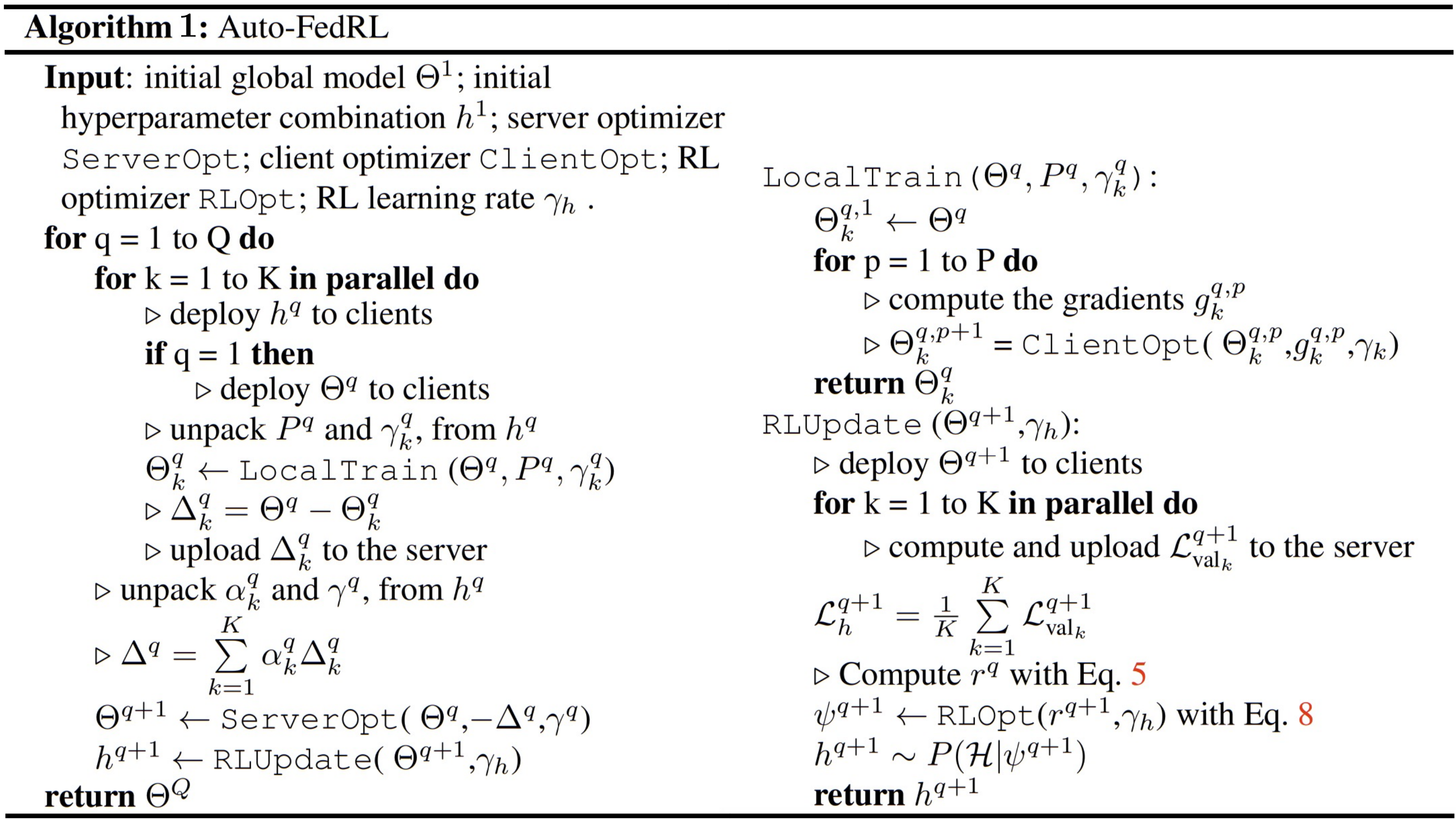}
\end{figure}
\begin{figure}[tbp]
\centering
\includegraphics[width=0.6\columnwidth]{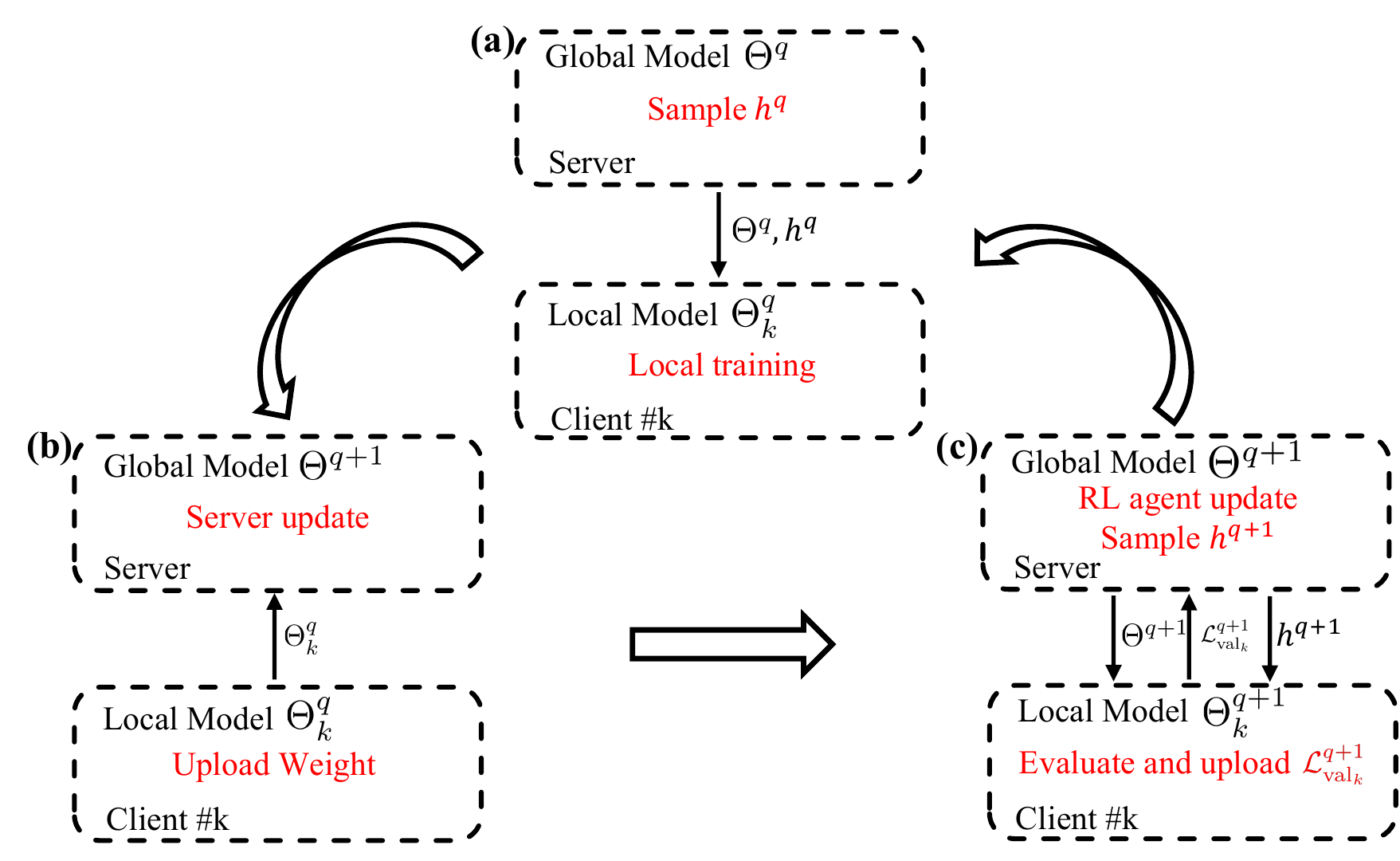}
\caption{The schematics of Auto-FedRL at round $q$.}\label{fig1}
\end{figure}

\noindent\textbf{Full Algorithm. } The overview of Auto-FedRL framework is presented in Alg.~\textcolor{red}{1} and Fig.~\ref{fig1}.
At each training round $q$, the training of Auto-FedRL consists of following steps: \textbf{(i)} As shown in Fig.~\ref{fig1}(a), clients receive the global model $\Theta^{q}$ and hyperparameters $h^{q}$. Clients perform {\tt LocalTrain} based on the received hyperparameters. \textbf{(ii)} The updated local models are then uploaded to the server as shown in Fig.~\ref{fig1}(b). Instead of performing the average aggregation, we use \emph{pseudo-gradient} $-\Delta^q$ in Eq.~\ref{eq:3} to carry out the server update with a searchable server learning rate. 
\textbf{(iii)} Clients evaluate the received the updated global model $\Theta^{q+1}$ and upload the validation loss $\mathcal{L}_{\text{val}_{k}}^{q+1}$ to the server. The server performs the RL update as shown in {\tt RLUpdate} of Alg.~\textcolor{red}{1}. Here, we consider the applicability in a real-world scenario, in which each client maintains its own validation data rather than relying on validation data being available on the server. Then, the server computes the reward $r^{q+1}$ as in Eq.~\ref{eq:4} and updates the RL agent ({\tt RLOpt}) as in Eq.~\ref{eq:7}. Finally, hyperparameters for the next training round $h^{q+1}$ can be sampled from the updated hyperparameter distribution $P(\mathcal{H}| \psi^{q+1})$. As shown in Fig.~\ref{fig1}(c), the proposed method requires one extra round of communication between clients and the server for $\mathcal{L}_{\text{val}_{k}}$. It is worth noting that the message size of $\mathcal{L}_{\text{val}_{k}}$ is negligible. Thus, this extra communication can still be considered practical under our targeted scenario in which all clients have a reliable connection (\ie, multi-institutional collaborations in cross-silo FL).

\subsection{Datasets and Implementation Details}
\noindent\textbf{CIFAR-10. } We simulate an environment in which the number of data points and label proportions are imbalanced across clients. Specifically, we partition the standard CIFAR-10 training set~\cite{cifar10} into 8 clients by sampling from a Dirichlet distribution ($\alpha=0.5$) as in \cite{fedma}. The original test set in CIFAR-10 is considered as the global test set used to measure performance. VGG-9~\cite{vgg} is used as the classification network. All models are trained using the following settings: Adam optimizer for RL; SGD optimizer for clients and the server; $\gamma_h$ of $1\times10^{-2}$; initial learning rate of $1\times10^{-2}$; maximum rounds of 100; initial local epochs of 20; batch size of 64.

\noindent\textbf{Multi-national COVID-19 Lesion Segmentation. } This dataset contains 3D computed tomography (CT) scans of COVID-19 infected patients collected from three medical centers\footnote[1] {\url{https://wiki.cancerimagingarchive.net/display/Public/CT+Images+in+COVID-19}}~\cite{data1,data2,autofedavg,data3}. We partition this dataset into three clients based on collection locations as following: 671 scans from China (Client I), 88 scans from Japan (Client II), and 186 scans from Italy (Client III). Each voxel containing a COVID-19 lesion was annotated by two expert radiologists. The training/validation/testing data splits are as follows: 447/112/112 (Client I), 30/29/29 (Client II), and 124/31/31 (Client III). The architecture of the segmentation network is 3D U-Net~\cite{3dunet}. All models are trained using the following settings: Adam optimizer for RL and clients; $\gamma_h$ of $1\times10^{-2}$; SGD optimizer for the server; initial learning rate of $1\times10^{-3}$; initial local iterations of 300; maximum rounds of 300; batch size of 16. 

\noindent\textbf{Multi-institutional Pancreas Segmentation. } Here, we utilize 3D CT scans from three public datasets, including 281 scans from the pancreas segmentation subset of the Medical Segmentation Decathlon~\cite{p1} as Client I, 82 scans from the Cancer Image Archive (TCIA) Pancreas-CT dataset~\cite{p2} as Client II, and 30 scans from Beyond the Cranial Vault (BTCV) Abdomen dataset~\cite{p3} as Client III. The training/validation/testing data splits are as follows: 95/93/93 (Client I), 28/27/27 (Client II), and 10/10/10 (Client III). All models are trained using the same network architecture and settings as COVID-19 lesion segmentation except that the maximum rounds are 50.

\section{Experiments and Results}
In this section, the effectiveness of our approach is first validated on a heterogeneous data split of the CIFAR-10 dataset (Sec.~\ref{sec:cifar}).  
Then, experiments are conducted on two multi-institutional medical image segmentation datasets (\ie, COVID-19 lesion segmentation and pancreas segmentation) to investigate the real-world potential of the proposed Auto-FedRL (Sec.~\ref{sec:realworld}). Finally, detailed comparisons between discrete and continuous search space, and the exploration of deep RL agents are provided (Sec.~\ref{sec:ab}).
We evaluate the performance of our method against the following popular FL methods:
FedAvg~\cite{fedavg} and FedProx~\cite{fedprox} as well as FL-based hyperparameter optimization
methods: Auto-FedAvg~\cite{autofedavg}, and Mostafa \etal~\cite{intel}.

\begin{table}[tbp]
	\setlength{\tabcolsep}{30pt}
	\scriptsize
	\centering
	\caption{CIFAR-10 classification results. \redbf{Bold} and \blueud{Underline} indicate the best and the second best performance, respectively.}\label{tb1}
\begin{tabular}{l|c}
\hline
Method           & Accuracy (\%) \\ \hline\hline
FedAvg~\cite{fedavg}           & 88.43        \\ 
FedProx~\cite{fedprox}          & 89.45        \\ 
Mostafa \etal~\cite{intel}    & 89.86        \\ 
Auto-FedAvg~\cite{autofedavg}      & 89.16        \\ \hline\hline
Auto-FedRL(DS)       & 90.70        \\ 
Auto-FedRL(CS) & \blueud{90.85}        \\
Auto-FedRL(MLP)       & \redbf{91.27}        \\ \hline\hline
Centralized      & 92.56        \\ \hline
\end{tabular}
\end{table}
\begin{table}[t!]
	\setlength{\tabcolsep}{5.0pt}
	\scriptsize
	\caption{The computational details of different search strategies under the same setting on CIFAR-10.}\label{tb5}
	\centering
\begin{tabular}{l|cc}
\hline
Search Space Type & Memory Usage & Running Time for Search \\ \hline\hline
Discrete    & 42.8 GB     & 8.246 s             \\
Continuous  & \redbf{3.00 GB}      & \redbf{0.012 s}             \\ 
Continuous MLP  & \blueud{3.13 GB}      & \blueud{0.019 s}             \\ \hline
\end{tabular}
\end{table}

\subsection{CIFAR-10}\label{sec:cifar}
\begin{figure}[t!]
	\centering
	\includegraphics[clip, trim=0cm 0cm 0cm 0.1cm,
	width=0.75\columnwidth]{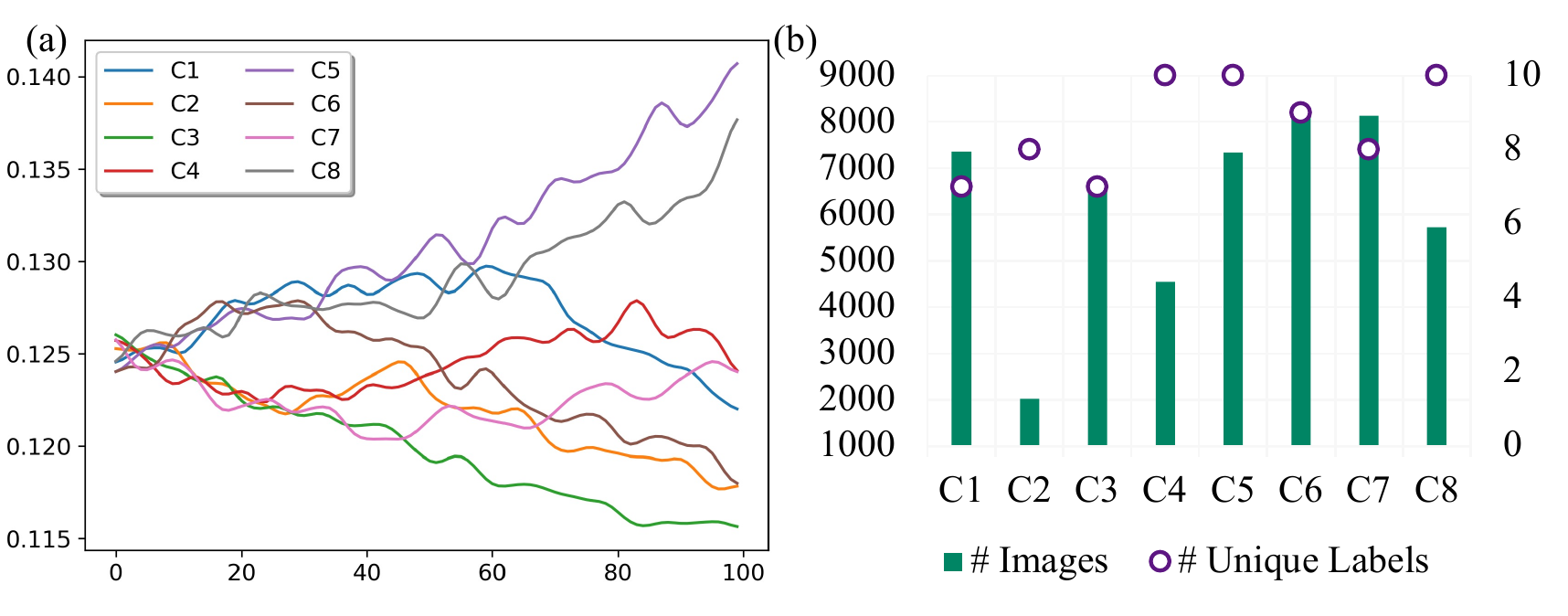}
	\caption{Analysis of the learning process of Auto-FedRL(MLP) in CIFAR-10. (a) the evolution of aggregation weights during the training. (b) the data statistics of different clients. }\label{fig4} 
\end{figure}

Table~\ref{tb1} shows the quantitative performance of different methods in terms of the average accuracy across 8 clients. We denote the model that is directly trained with all available data as \emph{Centralized} in Table~\ref{tb1}. We treat it as an upper bound when data can be shared.
As can be seen from this table, the proposed Auto-FedRL methods clearly outperform the other competing FL alternatives. Auto-FedRL(MLP) gains the best performance improvement by taking the advantage of a more complex RL agent design. To investigate the underlying hyperparameter change, we plot the evolution of aggregation weights in Fig.~\ref{fig4}. We found that the proposed RL agent is able to reveal more informative clients (i.e., clients containing more unique labels) and assign larger aggregation weights to those client's model updates. In particular, in Fig.~\ref{fig4}(a), C4 (red), C5 (purple), and C8 (gray) are gradually assigned three of the largest aggregation weights. As shown in Fig.~\ref{fig4}(b), although those three clients do not contain the largest number of images, all have the most number of unique labels (\ie 10 in CIFAR-10). This behavior further demonstrates the effectiveness of Auto-FedRL. Moreover, we provide the computational details of different search strategies to investigate their practicability under a same setting in Table~\ref{tb5}. Without losing performance, the proposed continuous search requires only 7\% memory usage but is 690$\times$ faster compared to the discrete search. While Auto-FedRL(MLP) introduces the deep RL agent, it is still 430$\times$ faster compared to the discrete version. Additional multi-dimensional comparisons~\cite{c3} are provided in the supplementary material. The notable computational efficiency and low memory usage of Auto-FedRL validate our motivation of making federated hyperparameter optimization more practical in real-world applications.

\begin{table}[tbp]
	\setlength{\tabcolsep}{12.0pt}
	\scriptsize
	\centering
		\caption{Multi-national COVID-19 lesion segmentation. $\dagger$ indicates significant improvement (p $\ll$ 0.05 in the Wilcoxon signed rank test) of the global model over the best counterpart.
}\label{tb2}
\begin{tabular}{l|cccc}
\hline
Method           & \multicolumn{1}{c|}{Client I} & \multicolumn{1}{c|}{Client II} & \multicolumn{1}{c|}{Client III} & \multicolumn{1}{c}{\textbf{Global Test Avg.}} \\ \hline\hline
Local only - I   & \multicolumn{1}{c|}{59.8}     & \multicolumn{1}{c|}{61.8}     & \multicolumn{1}{c|}{51.8}       & \multicolumn{1}{c}{57.8}            \\
Local only - II  & \multicolumn{1}{c|}{41.9}     & \multicolumn{1}{c|}{59.9}     & \multicolumn{1}{c|}{50.2}       & \multicolumn{1}{c}{50.7}            \\
Local only - III & \multicolumn{1}{c|}{34.5}     & \multicolumn{1}{c|}{52.5}     & \multicolumn{1}{c|}{65.9}       & \multicolumn{1}{c}{51.0}            \\ \hline\hline
FedAvg~\cite{fedavg}            & \multicolumn{1}{c|}{59.9}     & \multicolumn{1}{c|}{63.8}     & \multicolumn{1}{c|}{60.5}       & \multicolumn{1}{c}{61.4±0.2}      \\
FedProx~\cite{fedprox}           & \multicolumn{1}{c|}{60.3}     & \multicolumn{1}{c|}{64.9}     & \multicolumn{1}{c|}{60.5}       & \multicolumn{1}{c}{61.9±0.5}      \\
Mostafa \etal~\cite{intel}      & \multicolumn{1}{c|}{60.9}     & \multicolumn{1}{c|}{64.6}     & \multicolumn{1}{c|}{65.6}       & \multicolumn{1}{c}{63.7±0.3}       \\
Auto-FedAvg~\cite{autofedavg}       & \multicolumn{1}{c|}{60.3}     & \multicolumn{1}{c|}{65.3}     & \multicolumn{1}{c|}{64.8}       & \multicolumn{1}{c}{63.5±0.2}       \\ \hline\hline
Auto-FedRL(DS)       & \multicolumn{1}{c|}{59.3}     & \multicolumn{1}{c|}{65.6}     & \multicolumn{1}{c|}{68.9$^\dagger$}       & \multicolumn{1}{c}{\blueud{64.6}±0.2}                        \\
Auto-FedRL(CS)       & \multicolumn{1}{c|}{59.9}     & \multicolumn{1}{c|}{66.1$^\dagger$}     & \multicolumn{1}{c|}{68.2$^\dagger$}       & \multicolumn{1}{c}{\redbf{64.7}±0.1}        \\
Auto-FedRL(MLP)       & \multicolumn{1}{c|}{57.8}     & \multicolumn{1}{c|}{65.6}     & \multicolumn{1}{c|}{68.5$^\dagger$}       & \multicolumn{1}{c}{64.0±0.4} \\ \hline\hline
Centralized      & \multicolumn{1}{c|}{61.1}     & \multicolumn{1}{c|}{65.9}     & \multicolumn{1}{c|}{69.3}       & \multicolumn{1}{c}{65.4}                        \\ \hline
\end{tabular}
\end{table}
\begin{figure*}[t!]
	\centering
	\includegraphics[width=\textwidth]{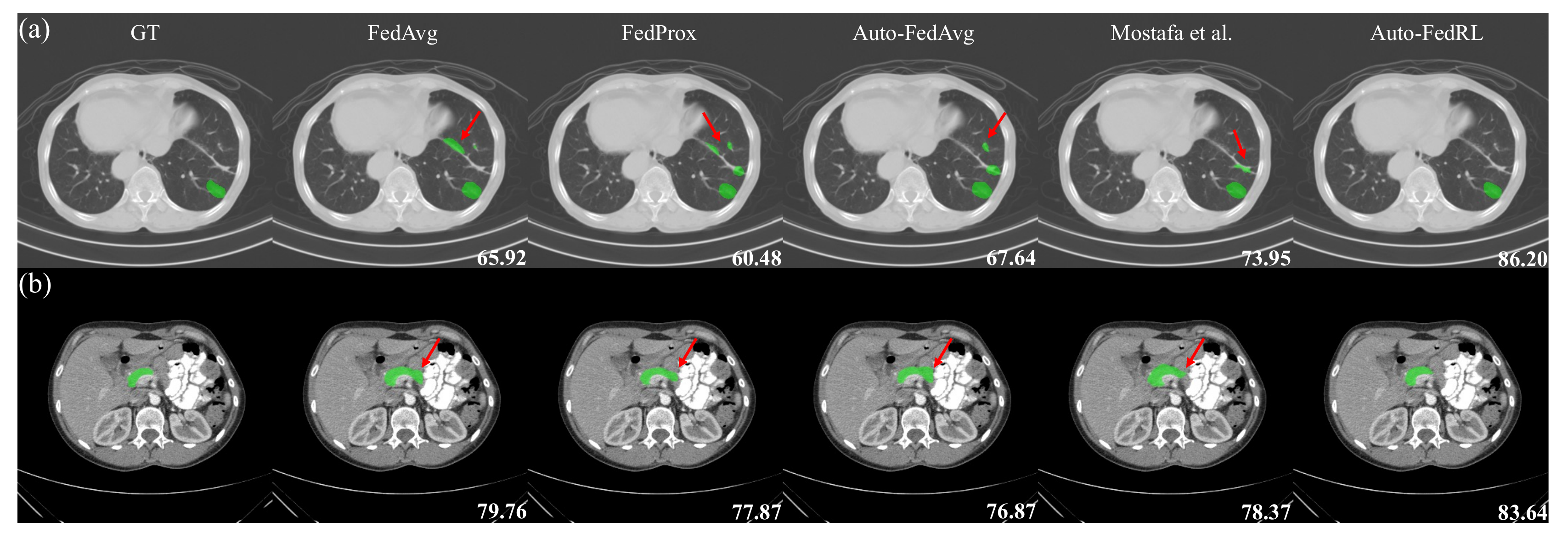}
	\caption{Qualitative results of different methods that correspond to (a) COVID-19 lesion segmentation of Client III and (b) Pancreas segmentation of Client II. GT shows human annotations in green and others show the segmentation results from different methods. Red arrows point to erroneous segmentation. The dice score is presented in the lower-right corner of each subplot.}\label{fig3}
\end{figure*}
\begin{figure*}[tbp]
	\centering
	\includegraphics[
	clip, trim=0cm 0.1cm 0cm 0cm,
	width=\textwidth]{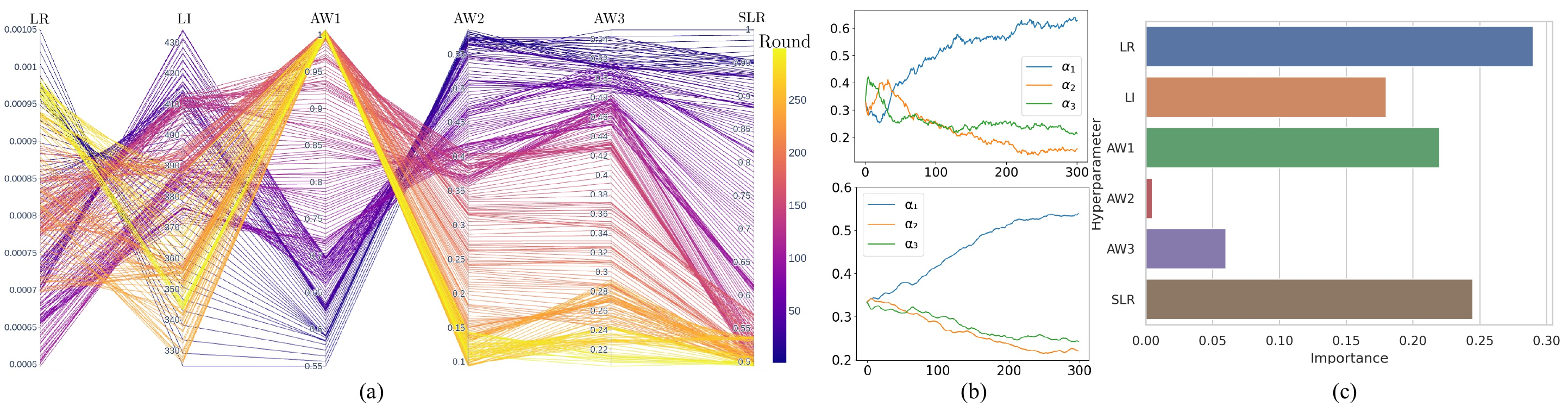}
	\caption{Analysis of the learning process of Auto-FedRL(CS) in COVID-19 lesion segmentation. (a) The parallel plot of the hyperparameter change during the training. LR, LI, AW, and SLR denote the learning rate, local iterations, the aggregation weight of each client, and the server learning rate, respectively. (b) The aggregation weights evolution of Auto-FedAvg in the \emph{top row} and Auto-FedRL(CS) in the \emph{bottom row}. (c) The importance analysis of different hyperparameters.}\label{fig5}
\end{figure*}

\subsection{Real-world FL Medical Image Segmentation}\label{sec:realworld}
\noindent\textbf{Multi-national COVID-19 Lesion Segmentation: } The quantitative results are presented in Table~\ref{tb2}. We show the segmentation results of different methods for qualitative analysis in Fig.~\ref{fig3}(a). Dice score is used to evaluate the quality of segmentation. We repeatedly run all FL algorithms 3 times and report the mean and standard deviation. The main metric of evaluating the generalizability of the global model is \emph{Global Test Avg.}, which is computed by the average performance of the global model across all clients. In the first three rows of Table~\ref{tb2}, we evaluate three locally trained models as the baseline. Due to the domain shift, all locally trained models exhibit low generalizability on the other clients. By leveraging the additional regularization on weight changes, FedProx (with the empirically best $\mu$=0.001) can slightly outperform the FedAvg baseline. Mostafa \etal that uses the RL agent to perform discrete search can achieve slightly better performance than Auto-FedAvg. We find that with the nearly constant memory usage and notable computational efficiency, the proposed Auto-FedRL(CS) achieves the best performance, outperforming the most competitive method~\cite{intel} by 1.0\% in terms of the global model performance, by 1.5\% and 2.6\% on clients II and III, respectively. The performance gap between the FL algorithm and centralized training is shrunk to only 0.7\%. Figure~\ref{fig5} presents the analysis of learning process in our best performing model. As shown in Fig.~\ref{fig5}(a), we can observe that the RL agent is able to naturally form the training scheduler for each hyperparameter (\eg, the learning rate decay for the client/server), which is aligned with the theoretical analysis about the convergence on non-iid data of FL algorithms~\cite{flconvergence}. Since Auto-FedAvg specially aims to learn the optimal aggregation weights, we compare the aggregation weights learning process between our approach and Auto-FedAvg in Fig.~\ref{fig5}(b). It can be observed that the two methods exhibit a similar trend of learning aggregation weights, which further demonstrates the effectiveness of Auto-FedRL in aggregation weights searching. Finally, we use FANOVA~\cite{FANOVA} to assess the hyperparameter importance. As shown in Fig.~\ref{fig5}(c), LR, SLR, and AW1 rank as the top-3 most important hyperparameters, which implies the necessity of tuning server-side hyperparameters in FL setting. 

\begin{table}[tbp]
	\setlength{\tabcolsep}{12.0pt}
	\scriptsize
	\centering
	\caption{Multi-institutional pancreas segmentation. $\dagger$ indicates significant improvement of the global model over the best counterpart.
}\label{tb3}
\begin{tabular}{l|c|c|c|c}
\hline
Method           & Clinet I           & Clinet II          & Clinet III           & \textbf{Global Test Avg.}    \\ \hline\hline
Local only - I   & 69.4                 & 71.4                 & 63.8                 & 68.2                \\
Local only - II  & 49.7                 & 75.5                 & 53.0                 & 59.3                \\
Local only - III & 42.4                 & 61.2                 & 51.1                 & 51.3                \\ \hline\hline
FedAvg~\cite{fedavg}            & 71.9                 & 78.4                 & 69.1                 & 73.1±0.3          \\
FedProx~\cite{fedprox}           & 72.0                 & 78.4                 & 69.6                 & 73.3±0.3          \\
Mostafa \etal~\cite{intel}       & \multicolumn{1}{c|}{74.4}     & \multicolumn{1}{c|}{79.4}     & \multicolumn{1}{c|}{72.1}       & \multicolumn{1}{c}{75.3±0.1}       \\
Auto-FedAvg~\cite{autofedavg}       & 71.3                 & 79.9                 & 71.5                 & 74.2±0.3          \\ \hline\hline
Auto-FedRL(DS)       & \multicolumn{1}{c|}{72.8}          & \multicolumn{1}{c|}{80.8$^\dagger$}          & \multicolumn{1}{c|}{74.7$^\dagger$}            & \multicolumn{1}{c}{76.1±0.4}                 \\
Auto-FedRL(CS) & \multicolumn{1}{c|}{73.0}          & \multicolumn{1}{c|}{82.2$^\dagger$}          & \multicolumn{1}{c|}{74.5$^\dagger$}            & \multicolumn{1}{c}{\blueud{76.5}±0.3}                 \\
Auto-FedRL(MLP) & \multicolumn{1}{c|}{73.2}          & \multicolumn{1}{c|}{81.2$^\dagger$}          & \multicolumn{1}{c|}{75.3$^\dagger$}            & \multicolumn{1}{c}{\redbf{76.6}±0.3}\\ \hline\hline
Centralized      & \multicolumn{1}{c|}{74.5}          & \multicolumn{1}{c|}{82.6}            & \multicolumn{1}{c|}{72.0}              & \multicolumn{1}{c}{76.3}                  \\ \hline
\end{tabular}
\end{table}
\noindent\textbf{Multi-institutional Pancreas Segmentation. } Table~\ref{tb3} and Fig.~\ref{fig3}(b) present the quantitative and qualitative results on this dataset, respectively. Similar to the results on COVID-19 segmentation, our Auto-FedRL algorithms achieves the significantly better overall performance. In particular, Auto-FedRL(MLP) outperforms the best counterpart by 1.3\%. We aslo observe that our methods exhibits better generalizability on the relatively smaller clients. Specifically, on Client III, Auto-FedRL(MLP) improves the Dice score from 51.1\% to 75.3\%, which is even 3.28\% higher than the centralized training. These results implies that by leveraging the dynamic hyperparameter tuning during the training, Auto-FedRL algorithms can achieve better generalization and are more robust towards the heterogeneous data distribution. As shown in Fig.~\ref{fig3}(b), the proposed methods have a better capacity of handling the challenging cases, which is consistent with the quantitative results. The detailed hyperparameter evolution analysis on pancreas segmentation is provided in the supplementary material.

\begin{table}[htbp]
	\setlength{\tabcolsep}{6.0pt}
	\scriptsize
	\centering
\caption{The search space ablation study on CIFAR-10.}\label{tb4}
\begin{tabular}{cccc|ccc}
\hline
\multicolumn{4}{c|}{Search Space}                 & \multicolumn{3}{c}{Search Strategy}    \\ \hline
LR         & LE         & AW         & SLR        & Discrete & Continuous & Continuous MLP \\ \hline\hline
\checkmark &            &            &            & 89.83    & \blueud{90.02}      & \redbf{90.12}          \\
\checkmark & \checkmark &            &            & 89.86    & \blueud{90.10}      & \redbf{90.49}          \\
\checkmark & \checkmark & \checkmark &            & 90.43    & \blueud{90.52}      & \redbf{90.87}          \\
\checkmark & \checkmark & \checkmark & \checkmark & 90.70    & \blueud{90.85}      & \redbf{91.27}          \\ \hline
\end{tabular}
\end{table}
\subsection{Ablation Study}~\label{sec:ab}
The effectiveness of the proposed continuous search and NN-based RL agent is demonstrated by the previous sets of experiments
in three datasets. Here, we conduct a detailed ablation study to analyze the benefit of individually adding each hyperparameter into the search space. As shown in Table~\ref{tb4}, the performance of trained global model can be further improved with the expansion of the search space, which also validates our motivation that the proper hyperparameter tuning is crucial for the success of FL algorithms. More visualizations, experimental results, and the theoretical analysis to guarantee the convergence of Auto-FedRL are provided in the supplementary material.

\section{Conclusion and Discussion}
In this work, we proposed an online RL-based federated hyperparameter optimization framework for realistic FL applications, which can dynamically tune the hyperparameters during a single trial, resulting in improved performance compared to several existing baselines. To make federated hyperparameter optimization more practical in real-world applications, we proposed Auto-FedRL(CS) and Auto-FedRL(MLP), which can operate on continuous search space, demand nearly constant memory and are computationally efficient. By integrating the adaptive federated optimization, Auto-FedRL supports a more flexible search space to tune a wide range of hyperparameters. The empirical results on three datasets with diverse characteristics reveal that the proposed method can train global models with better performance and generalization capabilities under heterogeneous data distributions.

While our proposed method yielded a competitive performance, there are potential areas for improvement. 
First, we are aware that the performance improvement brought by the proposed method is not uniform across participating clients. Since the proposed RL agent jointly optimizes the whole system, minimizing an aggregate loss can lead to potentially advantage or disadvantage a particular client's performance. We can also observe that all FL methods exhibit a relatively small improvement on the client with the largest amount of data. This is a common phenomenon of FL methods since the client itself already provides diverse and rich
training and testing data. Future research could include additional fairness constraints~\cite{fair1,fair2,fair3,fair4,fair5} to achieve a more uniform performance distribution across clients and reduce potential biases. 
Second, the NN-based RL agent could be benefiting from transfer learning. The effectiveness of RL transfer learning has been demonstrated in the literature for related tasks~\cite{tl}. Pre-training the NN-based agent on large-scale FL datasets and then finetuning on target tasks may further boost the performance of our approach.


\clearpage
%
%
\bibliographystyle{splncs04}

\end{document}